\documentclass[%
reprint,
 amsmath,amssymb,
aps,
pra,
]{revtex4-2}

\usepackage{graphicx}
\usepackage{siunitx}
\DeclareSIUnit\gauss{G}
\usepackage{physics}
\usepackage{ulem}
\usepackage{xcolor}
\usepackage{dcolumn}
\usepackage{bm}
\usepackage{hyperref}
\usepackage[mathlines]{lineno}

\begin{document}

\preprint{APS/123-QED}

\title{Measuring the environment of a Cs qubit with dynamical decoupling sequences}

\author{Sabrina Burgardt}
\author{Simon B. Jäger}
\author{Julian Feß}
\author{Silvia Hiebel}
\author{Imke Schneider}
\author{Artur Widera}
\email{widera@rptu.de}

  \affiliation{Department of Physics and Research Center OPTIMAS, University of Kaiserslautern-Landau, Erwin-Schrödinger-Straße 46, D-67663 Kaiserslautern, Germany}

\date{\today}

\begin{abstract}
We report the experimental implementation of dynamical decoupling on a small, non-interacting ensemble of up to 25 optically trapped, neutral Cs atoms.
The qubit consists of the two magnetic-insensitive Cs clock states $\ket{F=3, m_F=0}$ and $\ket{F=4, m_F=0}$, which are coupled by microwave radiation.
We observe a significant enhancement of the coherence time when employing Carr-Purcell-Meiboom-Gill (CPMG) dynamical decoupling.
A CPMG sequence with ten refocusing pulses increases the coherence time of $\SI{16.2(9)}{\milli\second}$  by more than one order of magnitude to $\SI{178(2)}{\milli\second}$. 
In addition, we make use of the filter function formalism and utilize the CPMG sequence to measure the background noise floor affecting the qubit coherence, finding a power-law noise spectrum $1/\omega^\alpha$ with $\mathit{\alpha} = \SI{0.89(2)}{}$.
This finding is in very good agreement with an independent measurement of the noise in the intensity of the trapping laser.
Moreover, the measured coherence evolutions also exhibit signatures of low-frequency noise originating at distinct frequencies. 
Our findings point toward noise spectroscopy of engineered atomic baths through single-atom dynamical decoupling in a system of individual Cs impurities immersed in an ultracold $^{87}$Rb bath. 
\end{abstract}

\maketitle

\section{\label{sec:introduction}Introduction}
The tremendous progress in manipulating and measuring ultracold quantum gases makes them a versatile platform for quantum technologies. 
Electric and magnetic field control enable the precise creation of quantum superpositions for targeted applications in quantum computing~\cite{Gyongyosi2019}, quantum simulation~\cite{Georgescu2014}, and quantum sensing~\cite{Degen:2017}.
In all these applications, information is stored in the relative phase of such a quantum superposition in so-called quantum bits (qubits).
These coherent superpositions, however, are usually very fragile because they are sensitive to external perturbations.
In particular, even in highly specialized labs, quantum systems cannot be perfectly isolated from their environment, which will lead to dissipation and eventually to decoherence, i.e., the decay of a pure quantum state into a statistical mixture. 
In this regard, the reduction of decoherence and the extension of coherence times is a prime challenge in quantum science which will enable the broad applications of quantum technologies.

One of the main sources of decoherence is dephasing caused by time-dependent (classical) fluctuations.
It is known that such dephasing can, in principle, be reversed by so-called dynamical decoupling, a coherent control-pulse method that was originally developed in
the field of solid-state physics.
The primary purpose of dynamical decoupling is the effective reduction of the coupling between the qubit and a mostly generic environment by a sequence of pulses. 
These single \cite{Hahn1950} and multi-pulse sequences \cite{CarrPurcell1954, MeiboomGill1958, Uhrig2007, Khodjasteh2007} have already been used to extend coherence in various physical systems, ranging from spin ensembles \cite{Du2009}, semiconductor quantum dots \cite{Barthel2010, Bluhm2011}, nitrogen-vacancy centers in diamond \cite{Lange2010, Ryan2010}, superconducting qubits \cite{Bylander2011}, ensembles of ultracold atoms and ions \cite{Biercuk2009Nature, Biercuk2009, Sagi2010} to the limit of single atoms and ions \cite{Szwer2010, Yu2013, Chow2021}. 
Consequently, many key characteristics and possible causes of qubit decoherence in ultracold atomic systems have already been discussed and explained in various experimental and theoretical works \cite{Gardiner2000, Kuhr2005, Windpassinger2008, Almog2011, Chuu2009, Gerasimov2021}.

Remarkably, such sequences cannot only be utilized to partially reverse the decoherence dynamics but also to perform noise spectroscopy, i.e., to measure details of the coupled, unknown degrees of freedom of the environment \cite{Bylander2011, Alvarez:2011, Yuge:2011, Cywinski2013, Szankowski2017, Krzywda2019, Chow2021}.
This first appears to be counter-intuitive since sensing is usually more effective if the coupling between the probe and the measurement object is strong. 
The idea of noise spectroscopy through dynamical decoupling, by contrast, is the modification of the spectral properties of the qubit which probes the environmental noise spectrum in different frequency ranges.
While one could expect that the measurement of all environmental effects is a hopeless task, the most relevant effects can be directly extracted from the detailed knowledge of the qubit's dynamics. 
This creates the possibility to recover details of the environment by controlling and measuring the qubit's time evolution.
From a fundamental point of view, the dynamical driving of the qubit can be used for the creation of correlated states between the environmental and the qubit's degrees of freedom, closely related to spin-boson~\cite{Leggett:1987}, polaron~\cite{Yan:2020, Skou2021}, and Kondo physics~\cite{Kondo1964, Hewson1993}. 
From a more practical point of view, the gained knowledge about the environment can be utilized to improve the experiment by applying optimized sequences tailored to the situation in the lab. 

In this paper, we exploit the dynamical decoupling method to determine essential properties of the environmental noise spectrum affecting the qubit coherence in a small ensemble of optically trapped, neutral $^{133}$Cs atoms.  
We use the two magnetic-insensitive Zeeman levels $\ket{F=3, m_F=0}$ and $\ket{F=4, m_F=0}$ of the $6 S_{1/2}$ ground state as qubit states and couple them by microwave radiation.
Such atoms have been immersed into an ultracold atomic gas recently \cite{Adam2022} for probing coherence and dephasing dynamics of individual Cs impurities beyond the effect of electromagnetic fields.
In this work, we demonstrate how dynamical qubit control can give access to essential properties of a coupled environment.
Our results represent a subsequent stepping stone towards utilizing the Cs qubit as a quantum probe for correlated environments consisting of interacting quantum particles.

The paper is structured as follows. 
Sec.~\ref{sec:experimental-system} comprises a description of the experimental setup and a detailed characterization of the qubit. 
In Sec.~\ref{sec:theory}, we introduce the theoretical description of the dynamically driven qubit in a decohering environment.
Then, in Sec.~\ref{sec:cpmg-measurements}, we  present the experimental realization of the dynamical decoupling sequence and a comparison to theory. 
After that, we conclude our results in Sec.~\ref{sec:conclusion-outlook} and provide a brief outlook. 

\section{\label{sec:experimental-system}Experimental system}
\subsection{\label{sec:experimental-setup}Experimental setup}
\begin{figure}[bp]
\begin{center}
\includegraphics[width=8.6 cm]{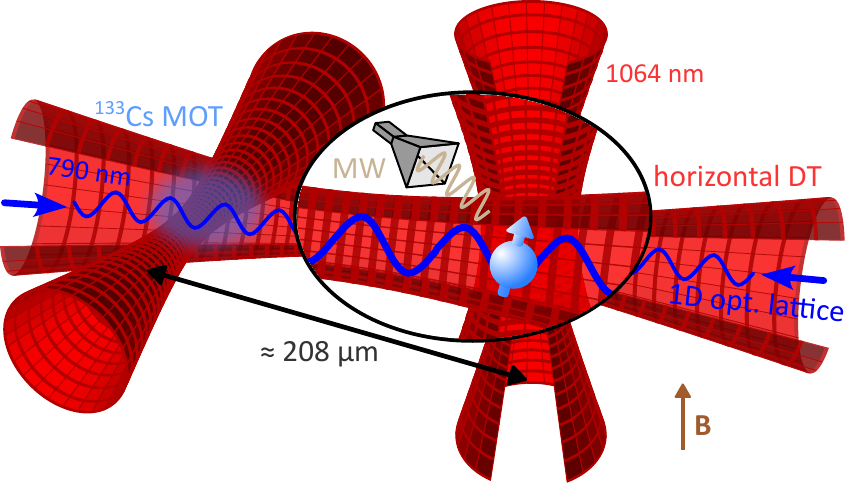}
\caption{\textbf{Sketch of the experimental setup.} Three running dipole trapping beams at a wavelength of $\SI{1064}{\nano \meter}$ (red) build two independent crossed optical dipole traps sharing one trapping beam along the horizontal direction. The one-dimensional optical lattice (dark blue) is a standing-wave dipole trap formed by a pair of laser beams at a wavelength of $\SI{790}{\nano \meter}$ superimposed with the shared horizontal dipole trap. Microwave (MW) radiation couples the Cs clock states.}
\label{fig:experimental-setup}
\end{center}
\end{figure}

A schematic overview of the trapping geometry in our experimental setup is given in Fig.~\ref{fig:experimental-setup}.
The preparation of the Cs atoms is a three-stage process.
First, about 50 Cs atoms are captured in a high-gradient magneto-optical trap and loaded into a crossed optical dipole trap formed by the intersection of two  $\SI{1064}{\nano \meter}$ laser beams.
Subsequently, the Cs atoms are further cooled to a temperature of $T_{\rm{Cs}} = \SI{5.9(3)}{\micro \kelvin}$ using degenerate Raman sideband cooling (see Refs.~\cite{Alt2003, Adam2022} for additional information on the temperature measurement).
In addition to the cooling effect, Raman cooling optically pumps the Cs atoms to the absolute ground state $\ket{F=3, m_F=3}$.
The Cs atoms are then transferred to the magnetic-insensitive state $\ket{F=3, m_F=0}$ by four successive microwave Landau-Zener sweeps \cite{Adam2022}. 
We remove residual Cs atoms that are not in $\ket{F=3, m_F=0}$ from the trap by a subsequent cleaning scheme, which is based on the spin selective readout in Ref.~\cite{Schmidt2018} and consists of a combination of microwave Landau-Zener sweeps and resonant laser pulses.
This preparation process eventually leaves about $\SIrange{15}{25}{}$ Cs atoms in the desired state $\ket{F=3, m_F=0}$ and, at most, two Cs atoms in states $\ket{F=3, m_F \neq 0}$ \cite{AtomNumberOffset}.
In the last step, a one-dimensional optical lattice formed by two counter-propagating $\SI{790}{\nano \meter}$ laser beams is used as a conveyor belt to transport the Cs atoms to a second crossed optical dipole trap located at an axial distance of about $\SI{208}{\micro \meter}$.
Here, a coherent superposition of the Cs clock states is prepared, as explained in the subsequent section below, and the dynamical decoupling sequence is applied.
Importantly, the Cs atoms are always trapped in the static optical lattice during the sequence. 
The lattice creates a repulsive potential for the Cs atoms along the horizontal direction and thereby freezes the atoms' position in this dimension. This results in trapping frequencies of $\mathit{\omega}_{\text{ax}} = 2 \mathrm{\pi} \times \SI{60.7}{\kilo \hertz}$ and $\mathit{\omega}_{\text{rad}} = 2 \mathrm{\pi} \times \SI{694}{\hertz}$ in the axial and the radial direction, respectively. 
Moreover, it ensures that Cs-Cs interactions are negligible since, at most, one Cs atom can be trapped in each lattice site.
At the end of each dynamical decoupling sequence, a resonant laser pulse removes all Cs atoms in $\ket{F=4}$  from the trap, and the optical lattice is used for spatially-resolved fluorescence imaging of Cs atoms in $\ket{F=3, m_F=0}$. 

\subsection{\label{sec:qubit-characterization}Qubit characterization}
We start with a detailed characterization of the qubit transition before providing information about the dynamical decoupling sequence in Sec.~\ref{sec:theory}.
For all measurements shown in the following, the external magnetic field at the atoms' position is calibrated to a value of $\SI{198.5}{\milli\gauss}$ (see Appendix~\ref{app:bfield-calibration} for further details on the magnetic field calibration).

\begin{figure}
\begin{center}
\includegraphics[width=8.6 cm]{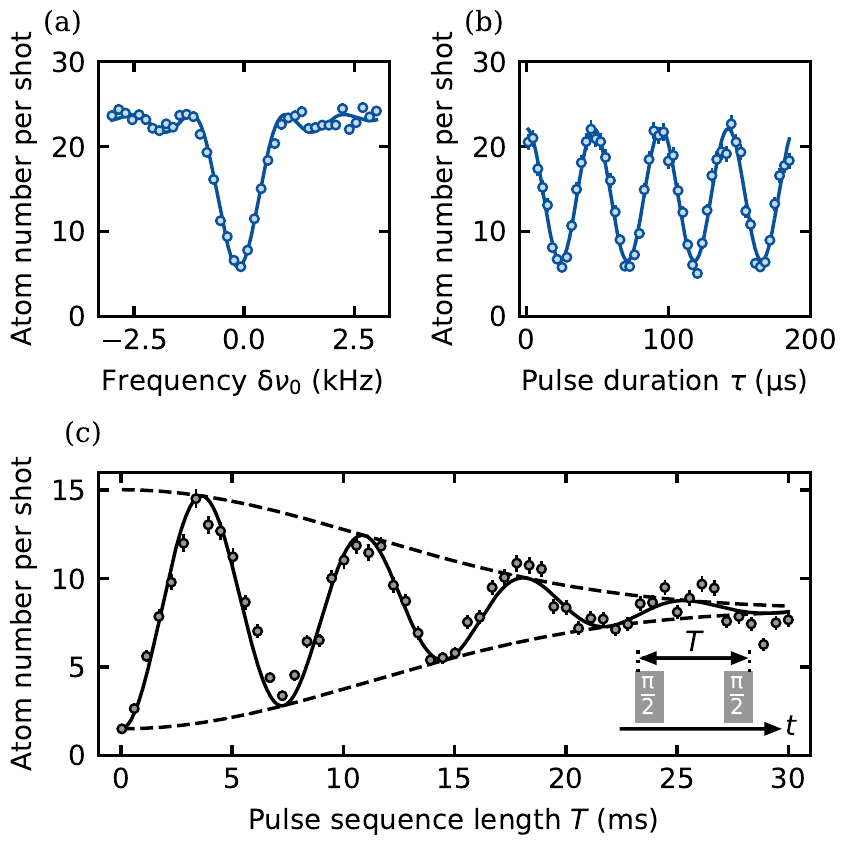}
\caption{\textbf{Qubit characterization.} (a)
Microwave spectroscopy of the qubit transition. The solid line shows a fit according to Eq.~(\ref{eq:mw-spectrum}). (b) Resonant Rabi oscillations for the qubit transition. (c) Ramsey fringes fitted with Eq.~(\ref{eq:fit-Ramsey-Titer}). Each data point in the panels (a) to (c) is an average of typically $\SIrange{50}{60}{}$ independent experimental runs. The error bars indicate the statistical uncertainties in the atom number determination.}
\label{fig:qubit-characterization}
\end{center}
\end{figure}

The qubit's transition frequency is measured using microwave spectroscopy, taking care to avoid power broadening.
The corresponding resonant Rabi frequency $\mathit{\Omega}_{\text{R}} = 2 \mathrm{\pi} \times \SI{619(2)}{\hertz}$ is independently measured, and Fig.~\ref{fig:qubit-characterization}(a) shows the spectrum of a microwave square pulse with $\tau = \SI{725}{\micro\second}$ duration. 
The signal follows the typical shape of the spectrum of a Fourier-limited square pulse 
\begin{equation}
N_\text{Cs}(\mathrm{\delta} \nu_{0}) = A \frac{\mathit{\Omega}_{\text{R}}^2}{\mathit{\Omega}^2} \sin^2 \left(\frac{\mathit{\Omega} \mathit{\tau}}{2} \right) + C,
\label{eq:mw-spectrum}
\end{equation}
where $\mathit{\Omega}^2 = 4 \mathrm{\pi}^2 (\mathrm{\delta} \nu_{0} - \mathrm{\delta} \nu_{0, \text{res}})^2 + \mathit{\Omega}_{\text{R}}^2$, and $A$  and $C$ are the amplitude and the offset of the fit, respectively.
The peak frequency is shifted from the exact atomic resonance frequency $\nu_0 = \SI{9.192 631 77}{\giga \hertz}$ by $\mathrm{\delta} \nu_{0, \text{res}}= \SI{-133(12)}{\hertz}$. 
The two main contributions to this shift are the differential light shift $\delta_{\text{DT}}$ of the trapping field and the quadratic Zeeman shift $\delta_B$. We find that the measured shift is in good agreement with the expected shift $\mathrm{\delta} \nu_{0, \text{theo}}= \SI{-123}{\hertz}$ calculated from experimental trapping parameters and the Breit-Rabi formula.

For dynamical decoupling sequences with multiple \mbox{($\pi$-)}pulses, the pulse durations are  important timescales determining the duration of the sequences and must be short compared to all other experimental timescales.
We achieve for all dynamical decoupling measurements a resonant Rabi frequency of $\mathit{\Omega}_{\text{R}} = 2 \mathrm{\pi} \times \SI{21.14(4)}{\kilo \hertz}$ [Fig.~\ref{fig:qubit-characterization}(b)], which corresponds to a $\pi$-pulse duration of $\tau_\pi = \SI{23.65}{\micro\second}$.
The Cs clock transition chosen is electric-dipole forbidden so that the excited state is long-lived, and $T_1$ decay occurs on the timescale of seconds and will therefore be neglected.  
Another important timescale is the lifetime of the Cs atoms in the optical lattice, which is limited by heating due to the relative phase noise between the two laser beams to about $\SI{685(81)}{\milli \second}$.   
Moreover, a sufficient timescale separation between coherence time and $\pi$-pulse duration is required for realizing multi-pulse 
sequences.
To measure the coherence time in the absence of dynamical decoupling, we apply a Ramsey sequence consisting of two microwave $\mathrm{\pi}/2$-pulses with variable sequence length $T$ [Fig.~\ref{fig:qubit-characterization}(c)]. 
The microwave frequency  is set to the unperturbed transition frequency $\nu_0$, i.e., detuned by approximately \SI{133}{\hertz} from the actual qubit resonance so that several fringes can be observed within the decay time. 
However, the pulses can still be approximated as near-resonant, as the detuning is much smaller than the bare Rabi frequency.
For such a sequence, the expected population dynamics can be described by
\begin{equation}
\label{eq:fit-Ramsey-Titer}
	N_\text{Cs}(T) = \frac{A}{2} \left[1 - \cos \left(2\mathrm{\pi} |\mathrm{\delta} \nu_{0, \text{res}}| T\right) e^{ -(T/\tilde{\mathit{\tau}}_\text{c})^2}  \right] + C,
\end{equation}
where $A$ and $C$ are the amplitude and offset of the fit. 
Previously, this functional behavior was a heuristic assumption that describes accurately the dynamics of the experiment. 
Later in this paper, in Sec.~\ref{sec:cpmg-measurements}, we will provide a justification of this formula which is a consequence of the measured noise spectrum.  
We fit Eq.~(\ref{eq:fit-Ramsey-Titer}) to our data in Fig.~\ref{fig:qubit-characterization}(c) and find a frequency shift of $|\mathrm{\delta} \nu_{0, \text{res}}|= \SI{136.8(8)}{\hertz}$, close to the frequency shift measured through microwave spectroscopy, and a coherence time of $\tilde{\mathit{\tau}}_\text{c} = \SI{15.7(7)}{\milli\second}$.
A detailed study on inhomogeneous and homogeneous dephasing effects causing decoherence in an ensemble of Cs atoms confined by a far-off-resonant standing-wave optical dipole trap can be found in Ref.~\cite{Kuhr2005}. 
Inhomogeneous dephasing is mainly caused by inhomogeneous differential light shifts originating from the thermal motion of the atoms in the trapping potential, which lead to slightly different resonance frequencies among the ensemble.
The effect of this dephasing mechanism can be reversed by applying a spin-echo sequence comprising an additional $\mathrm{\pi}$-pulse between the two Ramsey $\mathrm{\pi}/2$-pulses. 
Homogeneous dephasing effects, instead, lead to homogeneous, time-dependent fluctuations of the transition frequency among the ensemble. 
The main causes of homogeneous dephasing in our experimental setup are intensity fluctuations of the dipole trap laser and heating of the atoms (e.g., due to relative phase noise between the two lattice laser beams). 
Fluctuating magnetic fields are another source of broadening which is, however, far less pronounced because of the use of magnetic-field-insensitive states. 
Additional homogeneous dephasing sources, that play a minor role in our experimental setup, are the pointing instability of the dipole trap laser and imperfections of microwave pulses. 
Dephasing caused by homogeneous broadening cannot be simply reversed by the spin-echo sequence.
Here, we need to apply dynamical decoupling sequences, which extend the spin-echo scheme by using multiple $\mathrm{\pi}$-pulses, to compensate for noise originating at different frequencies. 
The theoretical background for this is presented in the following section.

\section{\label{sec:theory}Qubit coherence in a noisy environment}
 We use a semiclassical model to describe the dynamics of the driven qubit coupled to environmental noise~\cite{Cywinski2008, Biercuk2011}. 
 Hereby, we study the dynamics of the density matrix $\hat{\rho}$ of the qubit, which is governed by the von Neumann equation
\begin{align}
\frac{\partial\hat{\rho}}{\partial t}=\frac{1}{i\hbar}\left[\hat{H},\hat{\rho}\right].
\end{align}
The corresponding Hamiltonian is 
\begin{equation}
\label{eq:2-level-hamiltonian}
	\hat{H}=\frac{\hbar\mathit{\Omega}(t)}{2}\hat{\sigma}_x+\frac{\hbar \beta(t)}{2}\hat{\sigma}_z,
\end{equation}
and reported in the frame rotating with the qubit eigenfrequency. 
Here, we have introduced the Pauli matrices $\hat{\sigma}_i$ with $i=x,y,z$ of the two qubit states $\ket{F=3, m_F=0}$ and $\ket{F=4, m_F=0}$. 

The driving of this qubit is described by the coherent, time-dependent Rabi frequency $\mathit{\Omega}(t)$. 
The influence of the environment is encoded in fluctuations described by the stochastic process $\mathit{\beta}(t)$. 
We assume that this noise has a vanishing mean, $\langle\mathit{\beta}(t)\rangle=0$, and is determined by the time-correlation function $C(t-t^\prime)=\langle\mathit{\beta}(t)\mathit{\beta}(t^\prime)\rangle$. The expectation values are taken over noise realizations.
The stochastic fluctuations are characterized by the noise spectrum
\begin{align}
S(\omega)=\int_{-\infty}^\infty dt^\prime e^{-i\omega t^\prime} C(t^\prime).
\end{align}
In our description, we have neglected noise-driven spin flips, which are expected to be relevant only on timescales much longer than we describe in the following. 
Moreover, in this work, we will consider a spectral density of the form 
\begin{align}
S(\omega)=\frac{S_0}{|\omega|^\alpha}, \label{eq:spectral density}
\end{align}
with a positive real number $S_0>0$ and a positive real exponent~$\alpha>0$, describing an algebraic decay of the spectral density. 
This spectral density describes a variety of noise processes in various physical scenarios including $1/f$-noise.

\begin{figure}
\begin{center}
\includegraphics[width=8.6 cm]{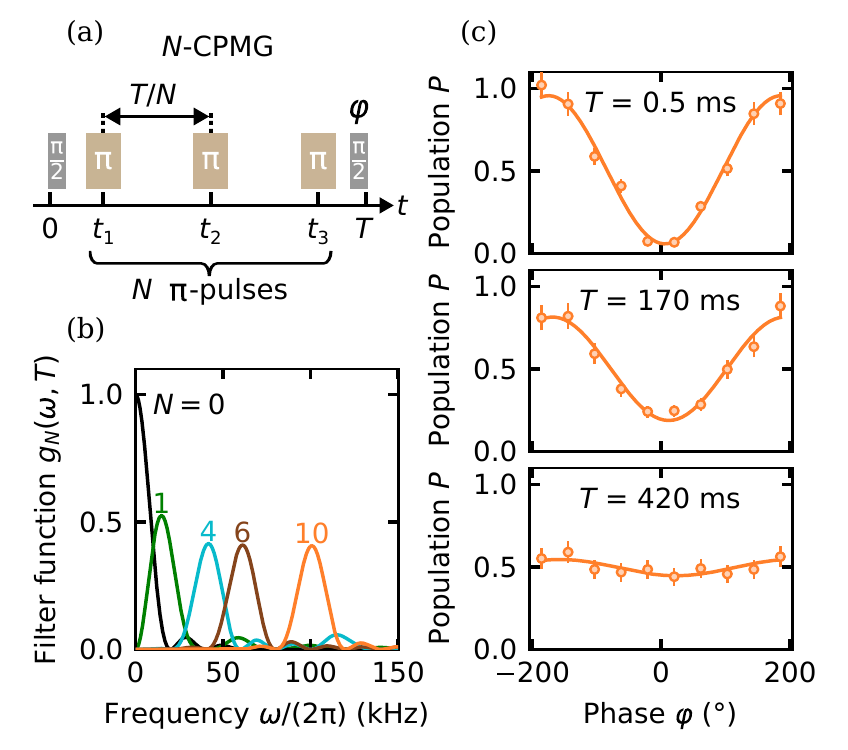}
\caption{$\bm{N}$-\textbf{CPMG dynamical decoupling.} (a) Schematics of the $N$-CPMG sequence for a total sequence length $T$. We scan the phase $\varphi$ of the second $\mathrm{\pi}/2$-pulse to measure the Ramsey fringes shown in panel (c). (b) $N$-CPMG filter functions $g_N(\omega, T)$ [Eq.~(\ref{eq:gpipulses})] for a fixed sequence length $T=\SI{50}{\milli\second}$. (c) Ramsey fringes for a fixed number of $N=10$ $\mathrm{\pi}$-pulses and various sequence lengths $T$ fitted with Eq.~\eqref{eq:Ramsey-fringe-fit}. The error bars are extracted from the statistical uncertainties in the atom number determination via standard error propagation.}
\label{fig:cpmg-theory}
\end{center}
\end{figure}

Similar to Ref.~\cite{Almog2011}, we derive a master equation which is valid in the weak coupling approximation. 
For this, we transform into a frame which is rotating with the driving field $\mathit{\Omega}(t)$, with $\hat{\varrho}=\hat{U}\hat{\rho}\hat{U}^\dag$ and $\hat{U}(t)=\exp(i\Phi(t)\hat{\sigma}_x/2)$, $\Phi(t)=\int_0^tdt^\prime\,\mathit{\Omega}(t^\prime)$. 
The dynamics of $\hat{\varrho}$ is then governed by the master equation
\begin{align}
\frac{\partial \hat{\varrho}(t)}{\partial t}=-\int_0^tdt^\prime\, \frac{C(t-t^\prime)}{4}\left\{\left[\hat{J}(t),\hat{J}(t^\prime)\hat{\varrho}(t)\right]+\mathrm{h.c.}\right\},\label{dyntilderho}
\end{align}
where we introduced $\hat{J}(t)=\hat{U}(t)\hat{\sigma}_z\hat{U}^\dag(t)$. 
This master equation can be mapped to the Bloch equations for the polarization ${\bf p}=(p_x,p_y,p_z)^T=\mathrm{Tr}[\boldsymbol{\sigma}\hat{\varrho}]$, with $\boldsymbol{\sigma}=(\hat{\sigma}_x,\hat{\sigma}_y,\hat{\sigma}_z)^T$. The Bloch equations are given by
\begin{align}
\frac{d{\bf p}}{dt}=-\int_0^t dt^\prime C(t-t^\prime){{\bf M}(t, t^\prime)}{\bf p}(t) \label{eq:Bloch}
\end{align}
with the matrix elements of ${\bf M}(t, t^\prime)$
\begin{align}
M_{xx}=&\cos[\Phi(t)]\cos[\Phi(t^\prime)]+\sin[\Phi(t)]\sin[\Phi(t^\prime)],\\
M_{yy}=&\cos[\Phi(t)]\cos[\Phi(t^\prime)],\\
M_{yz}=&-\cos[\Phi(t)]\sin[\Phi(t^\prime)],\\
M_{zy}=&-\sin[\Phi(t)]\cos[\Phi(t^\prime)],\\
M_{zz}=&\sin[\Phi(t)]\sin[\Phi(t^\prime)].
\end{align}
All other matrix elements are equal to zero. 
We will consider the case where $\mathit{\Omega}(t)$ is a series of $\delta$-shaped $\pi$-pulses applied at times $t_j$, described by $\mathit{\Omega}(t)=\sum_{j=1}^N\pi\delta(t-t_j)$. 
In this case, we find that $\sin(\Phi)=0$ for all times. 
The initial state in this protocol is obtained after applying a {$\pi/2$-pulse} to the ground state. 
Therefore, the initial state is polarized in the $y$ direction such that our initial conditions are $p_x(0)=0$, $p_y(0)=1$, $p_z(0)=0$. 
With these initial conditions, we can solve Eq.~\eqref{eq:Bloch} and find $p_x(T)=0$, $p_y(T)=V(T)$, $p_z(T)=0$, where $T$ is the total integration time and where we have introduced the visibility
\begin{align}
V(T)=e^{-\chi(T)}.
\end{align} The visibility is determined by the coherence integral 
\begin{equation}
\label{eq:coherence-intgeral}
	\chi(T) = \frac{T^2}{2\pi}  \int_0^\infty d\omega\, S(\omega) g(\omega, T) 
\end{equation}
and the filter function
\begin{align}
g(\omega,T)=&\frac{1}{T^2}\left|\int_0^T dt^\prime e^{i\omega t^\prime}\cos\left[\Phi(t^\prime)\right]\right|^2 .
\end{align}
For the sequence of $\pi$-pulses, we obtain
\begin{equation}
\label{eq:gpipulses}
\begin{split}
	g_N(\omega, T) = \frac{\left| 1 + (-1)^{1+N} e^{i \omega T} + 2 \sum_{j=1}^N (-1)^j e^{i \omega t_j} \right|^2}{(\omega T)^2}.
\end{split}
\end{equation}  
Until now, we have neither specified the instances $t_j$ for the {$\pi$-pulses} nor the total integration time $T$. 
In fact, the result given in Eq.~\eqref{eq:gpipulses} is general and can be used for an arbitrary sequence of $\delta$-shaped $\pi$-pulses, i.e., pulses with infinitely short pulse duration. 
In addition, one can also include a finite duration $\tau_\pi$ of the $\mathrm{\pi}$-pulses, which results in a slight modification of $g_N(\omega, T)$ that has been discussed in Refs.~\cite{Bylander2011, Biercuk2009, Biercuk2009Nature, Biercuk2011}.
In this paper, we employ the $N$-Carr-Purcell-Meiboom-Gill ($N$-CPMG) sequence, whose $\mathrm{\pi}$-pulses are applied at times $t_j = (2j-1)T/(2N)$~\cite{CarrPurcell1954, MeiboomGill1958}, and the $1$-CPMG sequence corresponds to the spin-echo sequence.
This pulse sequence consists of temporally equidistant $\pi$-pulses where the first pulse is applied at time ${t_1=T/(2N)}$ and the last pulse at ${t_N=T-T/(2N)}$ [see Fig.~\ref{fig:cpmg-theory}(a)].
We remark at this point that it has been shown that the filter function of the CPMG sequence is hardly modified by the effect of short $\mathrm{\pi}$-pulses with non-vanishing duration, as used in our experiment ($\tau_\pi / (T/N) < 10^{-1}$)~\cite{Biercuk2011}.
This justifies that we can neglect the influence of the finite pulse duration, which we do throughout this paper. 
The choice of the CPMG sequence enables a very efficient decoupling from noise with a spectrum $S(\omega)$ given by Eq.~\eqref{eq:spectral density}, especially if we consider no high-frequency cut-off~\cite{Biercuk2009Nature,Uhrig:2008}.
This results from the filter function $g_N(\omega, T)$, which is shown in Fig.~\ref{fig:cpmg-theory}(b) for different $N$-CPMG sequences and a fixed sequence length $T$. 
The filter functions' peaks shift to higher frequencies as the number of $\mathrm{\pi}$-pulses is increased, resulting in different frequency ranges of the noise spectrum contributing to decoherence [Eq.~(\ref{eq:coherence-intgeral})]. Consequently, since the spectral density $S(\omega)$ [Eq.~\eqref{eq:spectral density}] is decreasing with $\omega$, we obtain a lower total decoherence for an increasing number $N$ of $\pi$-pulses.
We mention, however, that we could also have chosen another pulse sequence such as the Uhrig sequence~\cite{Uhrig2007} which would show a similar performance in reducing the decoherence.

Besides the decoupling from noise, this band-pass filter property of $g_N(\omega, T)$ can also be exploited to perform noise spectroscopy.  
Here, we vary the number $N$ of $\mathrm{\pi}$-pulses and the sequence length $T$ such that the filter function samples the environmental noise spectrum $S(\omega)$. 
From the measured decoherence signal, we can then reconstruct the spectrum. 
To demonstrate this, we first use that the sum in Eq.~\eqref{eq:gpipulses} for the $N$-CPMG sequences can be written with the help of a geometric series
\begin{align}
g_N(u)=\frac{\left|1+(-1)^{1+N}e^{iu}+2e^{iu/(2N)}\dfrac{(-1)^Ne^{iu}-1}{e^{iu/N}+1}\right|^2}{u^2}.\label{eq:gnu}
\end{align}
Here, we have used that the expression of $g_N(\omega,T)$ for the $N$-CPMG sequence only depends on $u=\omega T$. 
The expression~\eqref{eq:gnu} exhibits for large $N$ a sharp peak whenever $u=(2k+1)N\pi$, where the denominator in the absolute value vanishes. 
For large $N$, we can approximate the filter function as an infinite series 
\begin{align}
g_N(u) \approx \sum_{k=0}^\infty\frac{4}{\left[(2k+1)\pi\right]^2}\mathrm{sinc}^2\left(\frac{u-(2k+1)N\pi}{2}\right), \label{eq:sumsinc}
\end{align}
with the introduction of $\mathrm{sinc}(x)=\sin(x)/x$. 
Since the slope of the spectral density Eq.~\eqref{eq:spectral density} is decreasing, and the appearing $\mathrm{sinc}$ functions in Eq.~\eqref{eq:sumsinc} evaluate the noise spectrum at larger and larger frequencies, we can exchange in Eq.~\eqref{eq:sumsinc} $\mathrm{sinc}^2(x)\approx\pi\delta(x)$ in the large $N$ limit. 
With this approximation, we can derive from Eq.~\eqref{eq:coherence-intgeral} the following formula
\begin{align}
\chi_N(T) \approx\left(\frac{T}{\tau_\text{c}}\right)^{1+\alpha},\label{eq:chiNanalytical}
\end{align}
where we have defined the coherence time
\begin{align}
\tau_\text{c}= N^{\alpha/(1+\alpha)} \tau_1,\label{eq:coherencetimeanalytic}
\end{align}
using
\begin{align}
\tau_1=\left\{4S_0\pi^{-2-\alpha}[1-2^{-2-\alpha}]\zeta(2+\alpha)\right\}^{-1/(1+\alpha)}, 
\end{align}
and the Riemann $\zeta$ function $\zeta(s)=\sum_{n=1}^\infty 1/n^s$. This finding is in agreement with Ref.~\cite{Cywinski2008}.

Equation~\eqref{eq:chiNanalytical} shows that the exponent $\alpha$ can directly be inferred from the visibility $V(T)$~\cite{Yuge:2011,Alvarez:2011,Medford:2012}. 
The latter exhibits an exponential behavior where the exponent of the argument is $1+\alpha$, determined by the algebraic behavior of the spectral density. 
Moreover, the coherence time $\tau_\text{c}$ [Eq.~\eqref{eq:coherencetimeanalytic}] is algebraically growing with the number $N$ of pulses.  
This increasing coherence time can be understood by the shift of the filter function's peak towards higher frequencies for an increasing number $N$ of $\mathrm{\pi}$-pulses [see Eq.~\eqref{eq:sumsinc} and Fig.~\ref{fig:cpmg-theory}(b)]. 
The latter leads to a reduction in the net integrated noise and thereby to an enhancement of the coherence time.
The exponent of the growth is $\alpha/(1+\alpha)$ which is monotonically increasing with increasing $\alpha$. 
Therefore, the noise reduction is more pronounced if the noise spectrum decays faster. For $\alpha\to\infty$, the coherence time is increasing linearly with $N$.

\section{\label{sec:cpmg-measurements} \textit{N}-CPMG measurements}
In the experiment, the effect of a $N$-CPMG sequence on the coherence time is investigated by tracking the visibility of Ramsey fringes for different sequence lengths $T$. 
To measure a Ramsey fringe for fixed $T$, we scan the phase $\varphi$ of the second $\mathrm{\pi}/2$-pulse from $\SIrange{-185}{185}{\degree}$ in 10 steps and measure the Cs population in $\ket{F=3, m_F=0}$ [Figs.~\ref{fig:cpmg-theory}(a) and \ref{fig:cpmg-theory}(c)].
We fit a sinusoidal function \cite{Adam2022}
\begin{equation}
      P(T, \varphi) = a \sin^2 \left(- \frac{\mathrm{\pi}}{\SI{360}{\degree}} \varphi + \frac{\mathrm{\pi}}{\SI{360}{\degree}} \Phi \right) + c
\label{eq:Ramsey-fringe-fit}
\end{equation}
to the data, where $a$, $c$ and $\Phi$ are the amplitude, the offset, and the free phase of the fit, respectively.
The fringe visibility is defined as 
\begin{equation}
V(T) = \frac{P_\text{max}-P_\text{min}}{P_\text{max} + P_\text{min}} = \frac{a}{a + 2c}.
\label{eq:Ramsey-fringe-visibility}
\end{equation}

\begin{figure}[b]
\begin{center}
\includegraphics[width=8.6 cm]{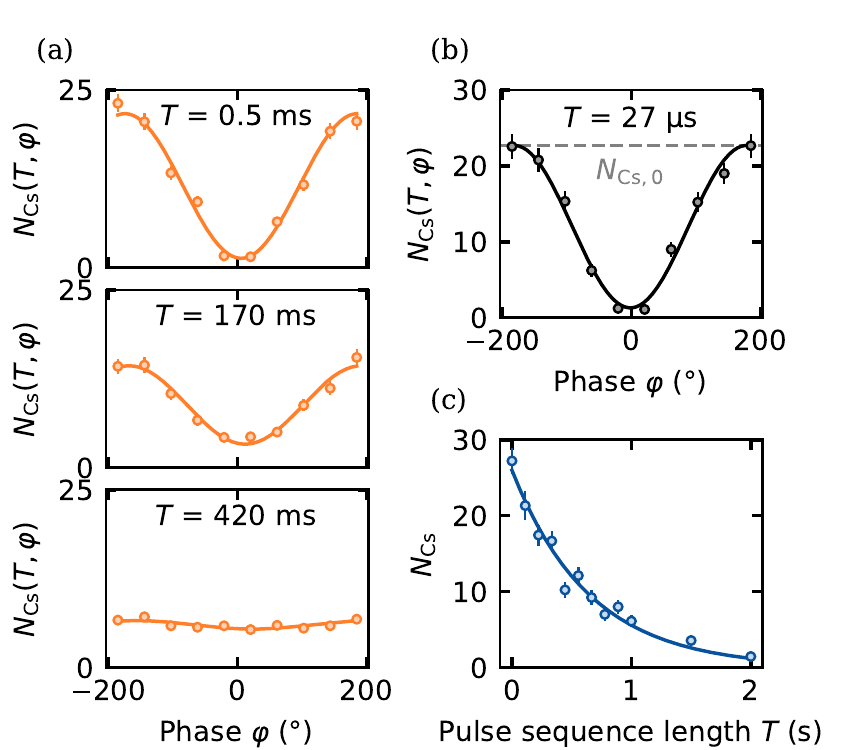}
\caption{\textbf{Ramsey fringe normalization.} (a) Ramsey fringes with total Cs atom number for a fixed number of $N=10$ $\mathrm{\pi}$-pulses and various sequence lengths $T$, as shown in Fig.~\ref{fig:cpmg-theory}(c) of the main text. (b) Ramsey fringe for $N=0$ $\mathrm{\pi}$-pulses and shortest pulse sequence length $T=\SI{27}{\micro\second}$. The maximal number of Cs atoms contributing to the signal $N_{\text{Cs},0}$ is obtained from a fit according to Eq.~(\ref{eq:Ramsey-fringe-fit}) (black solid line). (c) Lifetime in the optical lattice. The solid line shows an exponential fit for the extraction of the lifetime $\tau_{\scriptscriptstyle\text{LT}}$. }
\label{fig:fringe-normalization}
\end{center}
\end{figure}

The Ramsey fringes with total Cs atom number [Fig.~\ref{fig:fringe-normalization}(a)] need to be normalized to calculate the fringe visibility according to Eq.~(\ref{eq:Ramsey-fringe-visibility}).
We rescale the data for a given number $N$ of $\mathrm{\pi}$-pulses and pulse sequence length $T$ according to
\begin{equation}
\label{eq:fringe-normalization}
	P(T, \varphi) = \frac{N_\text{Cs}(T, \varphi)}{N_{\text{Cs}, \text{tot}}(T)},
\end{equation}
where the total number of Cs atoms contributing to the signal $N_{\text{Cs}, \text{tot}}(T)$ depends on $T$ due to the finite lifetime of the atoms in the optical lattice.
For each data set with a fixed number $N$ of $\pi$-pulses, the total atom number $N_{\text{Cs}, \text{tot}}(T)$ is deduced from two additional measurements.
First, we extract the maximal number of Cs atoms contributing to the signal $N_{\text{Cs},0}$ from a Ramsey fringe with $N=0$ $\mathrm{\pi}$-pulses and shortest possible pulse sequence length $T=\SI{27}{\micro\second}$ [Fig.~\ref{fig:fringe-normalization}(b)]. 
Second, we measure the lifetime  $\mathit{\tau}_{\scriptscriptstyle\text{LT}}$ of the Cs atoms in the optical lattice by monitoring the Cs atom number as a function of the pulse sequence length $T$ without applying microwave radiation [Fig.~\ref{fig:fringe-normalization}(c)].
We find typical lifetimes between $\mathit{\tau}_{\scriptscriptstyle\text{LT}} = \SI{539(37)}{\milli \second}$ and $\mathit{\tau}_{\scriptscriptstyle\text{LT}} = \SI{831(72)}{\milli \second}$.
The total atom number, which sets the scaling factor for the normalization, is then given by
\begin{equation}
\label{eq:tot-atom-number}
	N_{\text{Cs}, \text{tot}}(T) = N_{\text{Cs},0} \exp(-T / \mathit{\tau}_{\scriptscriptstyle\text{LT}}).
\end{equation}

\begin{figure*}
\begin{center}
\includegraphics[width=17.2 cm]{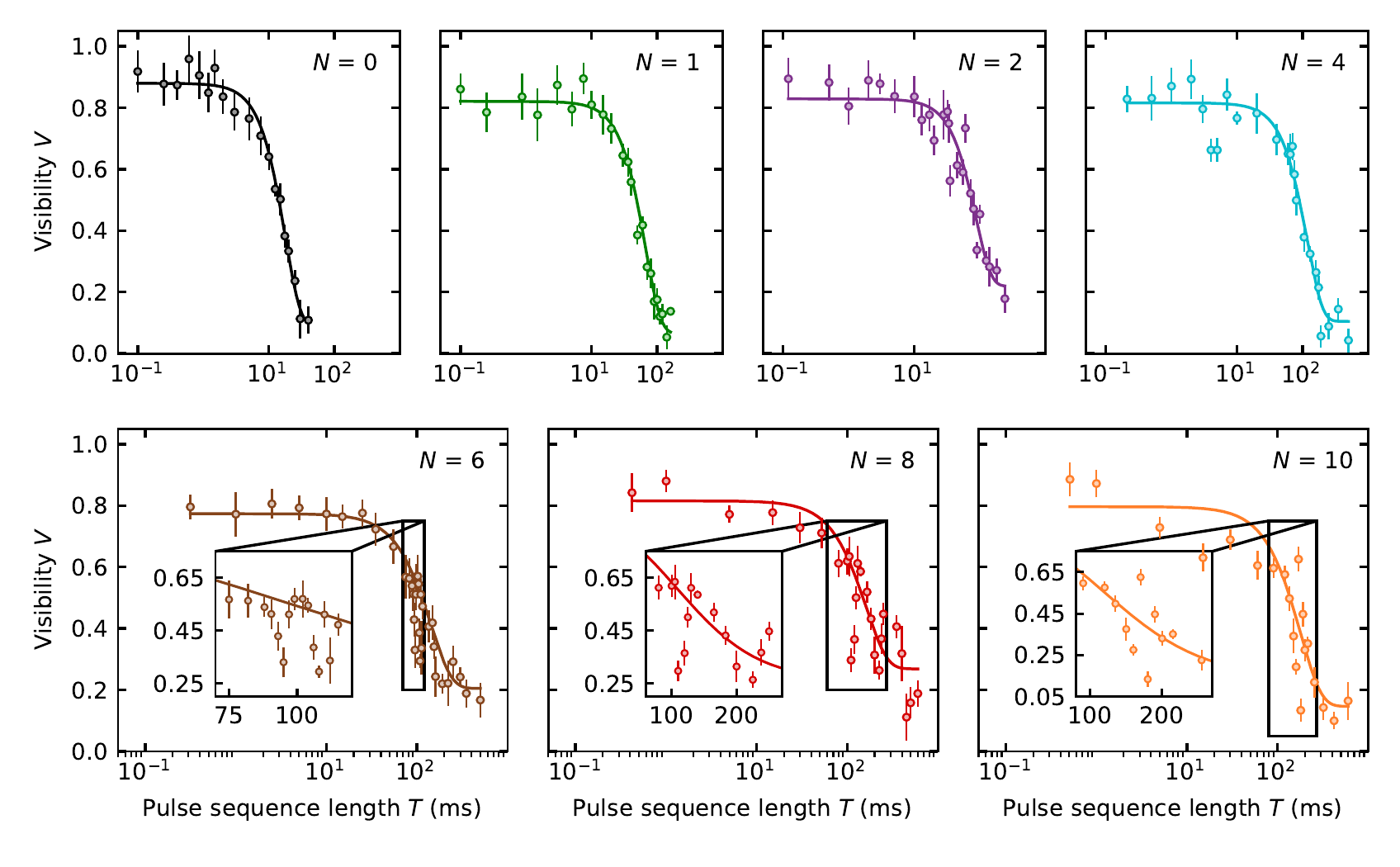}
\caption{\textbf{Visibility evolutions for $\bm{N}$-CPMG dynamical decoupling.} The solid lines show the theoretically predicted visibility evolution according to Eq.~(\ref{eq:Ramsey-fringe-visibility-fit}) with the fitted power-law noise spectrum $S(\mathit{\omega}) = S_0 / \mathit{\omega}^{\alpha}$ with $\mathit{\alpha} = \SI{0.89(2)}{}$ and $S_0=\SI{1288(122)}{\second}^{1-\alpha}$. The error bars result from the fitting uncertainty of each Ramsey fringe fit.}
\label{fig:cpmg-measurements}
\end{center}
\end{figure*}

Each data point in Fig.~\ref{fig:cpmg-theory}(c) [\ref{fig:fringe-normalization}(a)] is an average of typically $\SIrange{12}{15}{}$ experimental runs resulting in a total measurement time of about $\SIrange{18}{24}{\hour}$ for each $N$-CPMG sequence.
Figure~\ref{fig:cpmg-measurements} shows the measured visibility evolution for different numbers of $\mathrm{\pi}$-pulses ranging from $N=0$ to $N=10$. 
The data points predict, in general, a decaying dynamics of the visibility, which is slower for a larger number $N$ of $\pi$-pulses. 
To get a more quantitative access to this data, we fit 
\begin{equation}
V(T) = V_0 \: e^{-\chi_N (T)} + b.
\label{eq:Ramsey-fringe-visibility-fit}
\end{equation}
to the measured visibility. 
For the calculation of $\chi_N$, we calculate the coherence integral in Eq.~\eqref{eq:coherence-intgeral} with the expected form of the spectral density in Eq.~\eqref{eq:spectral density} and the filter function given by Eq.~\eqref{eq:gpipulses} for every number $N$ of $\pi$-pulses. This assures that we also take into account small-$N$ effects which are not present in the analytical result shown in Eq.~\eqref{eq:chiNanalytical}. Importantly, each $N$-CPMG data set owns its specific amplitude $V_0$ and offset $b$. 
In contrast, the parameters $S_0$ and $\alpha$ are global fitting parameters shared among all $N$-CPMG data sets. 
The fits are visible as the solid lines in Fig.~\ref{fig:cpmg-measurements} and agree well with the data points. 
From our fit, we find an exponent $\mathit{\alpha} = \SI{0.89(2)}{}$ and the value $S_0=\SI{1288(122)}{\second}^{1-\alpha}$. 

\begin{figure}
\begin{center}
\includegraphics[width=8.6 cm]{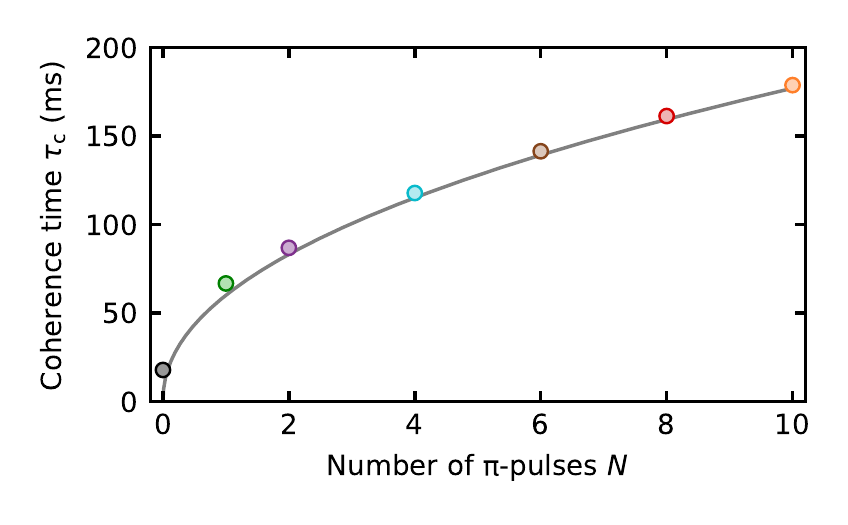}
\caption{\textbf{Coherence time as a function of the number
of $\bm{\mathrm{\pi}}$-pulses.} The solid line shows the analytical estimate of the coherence time according to Eq.~(\ref{eq:coherencetimeanalytic}), taking into account the fitted values $\mathit{\alpha} = \SI{0.89(2)}{}$ and $S_0=\SI{1288(122)}{\second}^{1-\alpha}$.}
\label{fig:coherencetime-numpulses}
\end{center}
\end{figure}

In addition, we can also extract the coherence time $\tau_\text{c}$, which is defined as $\exp[-\chi_N(\tau_\text{c})]=1/e$, from these fits. 
The values $\tau_\text{c}$ are shown in Fig.~\ref{fig:coherencetime-numpulses} as data points in a color scheme that matches the one used for the visibility data shown in Fig.~\ref{fig:cpmg-measurements} ($N=0$ black, $N=1$ green, etc.).
In the absence of dynamical decoupling ($N=0$), we find a coherence time of $\mathit{\tau}_\text{c} = \SI{16.2(9)}{\milli\second}$.
This value is in agreement with the coherence time of $\tilde{\tau}_\text{c} = \SI{15.7(7)}{\milli\second}$ obtained by fitting the population dynamics in Fig.~\ref{fig:qubit-characterization}(c) with Eq.~\eqref{eq:fit-Ramsey-Titer}. 
We want to underline that we have assumed a heuristic decay with $\exp\left[-(T/\tilde{\tau}_{\mathrm{c}})^2\right]$ in Eq.~\eqref{eq:fit-Ramsey-Titer}.
To validate this assumption, we can now compare this decay to the analytical result in Eq.~\eqref{eq:chiNanalytical} which predicts a decay with $\exp\left[-(T/\tau_\text{c})^{1+\alpha}\right] \approx \exp\left[-(T/\tau_\text{c})^{1.89}\right]$.
Consequently, the accurate result obtained from Eq.~\eqref{eq:fit-Ramsey-Titer} can be explained by the exponent $\alpha=0.89(2)$ which is close to one. For a growing number $N$ of $\mathrm{\pi}$-pulses, we obtain an increasing coherence time.  
In particular, we find for the maximum value of $N=10$ a coherence time of $\tau_\text{c}=\SI{178(2)}{\milli\second}$, which corresponds to an enhancement of more than one order of magnitude when compared to the $N=0$ case. 
To compare these results with analytical estimates, we have calculated Eq.~\eqref{eq:coherencetimeanalytic} from the fitted value of $\alpha$ and $S_0$. 
The result is shown as solid gray line in Fig.~\ref{fig:coherencetime-numpulses} and agrees better for larger numbers of $\mathrm{\pi}$-pulses. 
The analytical estimate predicts an algebraic growth with $N^{\alpha/(1+\alpha)}\approx N^{0.47}$. Compared to the theory, the experimental results also suggest that the observed coherence time can be further extended by adding more $\mathrm{\pi}$-pulses to the CPMG sequence since the Ramsey fringe visibility doesn't drop significantly. 
This also implies that imperfections in the $\mathrm{\pi}$-pulses play a minor role in our system.
Our analytical model in Eq.~\eqref{eq:coherencetimeanalytic} predicts an unbounded growth of the coherence time, which is a consequence of neglecting the natural $T_1$ decay time of the qubit states.
Dynamical decoupling sequences can, in principle, enhance the total coherence time only up to the $\tau_\text{c} = 2 T_1$ limit if energy relaxation is taken into account~\cite{Bylander2011}.
In our experiment, energy relaxation occurs on the timescale of seconds, but the maximum number of $\mathrm{\pi}$-pulses is eventually limited by the experimentally achievable minimum pulse duration and maximum sequence length. 
The former is determined by the available Rabi frequencies, and the latter is constrained by the lifetime of the Cs atoms in the optical lattice of about $\SI{685(81)}{\milli \second}$.
Consequently, we can never reach the physical limit $\tau_\text{c} = 2 T_1$ and expect the coherence time to grow for all experimentally accessible $N$-CPMG sequences.

In contrast to the strictly monotonous decaying fits visible in Fig.~\ref{fig:cpmg-measurements}, our experimental data also show narrow dips in the visibility for $N \geq 6$. These dips are highlighted in the insets of Fig.~\ref{fig:cpmg-measurements}
for $N=6,8,10$. We expect that these dips originate from several resonances in the noise spectrum. In fact, already a single resonance gives rise to the appearance of multiple dips in the visibility. This can be seen by considering a spectral density
\begin{align}
S(\omega)=S_1e^{-(\omega-\omega_0)^2/(2\Delta\omega^2)} \label{eq:Gauss}
\end{align}
which describes phenomenologically a resonance at $\omega_0$ with a finite width of $\Delta\omega$. 
Using Eq.~\eqref{eq:sumsinc} and Eq.~\eqref{eq:Gauss} in Eq.~\eqref{eq:coherence-intgeral}, we expect multiple dips at times $T_k(N)=(2k+1)N\pi/\omega_0$ for $k=0,1,2,\dots$. 
This makes it very difficult to determine from the experimental data whether the dips result from different resonances or from a single resonance. 
In particular, we could assume that the two dips in Fig.~\ref{fig:cpmg-measurements}
for $N=8$ found at $T_a$ and $T_b$ with $T_a\approx\SI{120}{\milli\second}$ and
$T_b\approx \SI{200}{\milli\second}$ originate from the same resonance $\omega_0$. 
In addition, if they correspond to neighboring dips, we can assume $T_b/T_a=(2k+3)/(2k+1)$ which can be compared to the data, $T_b/T_a\approx1.67$. 
This suggests that $k=1$ is a good fit with $T_b=T_{2}(N=8)$ and $T_a=T_{1}(N=8)$. 
From this, we can estimate the frequency which is approximately $\omega_0\approx 2\mathrm{\pi}\times \SI{100}{\hertz}$. 

\begin{figure}
\begin{center}
\includegraphics[width=8.6 cm]{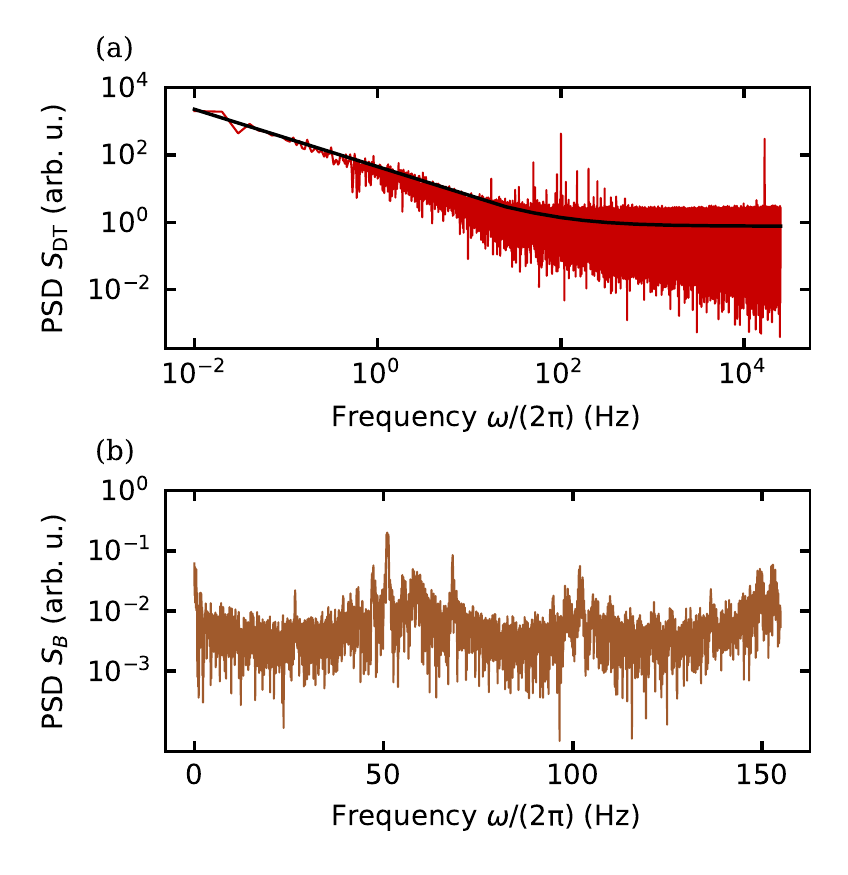}
\caption{\textbf{Environmental noise characterization} (a) PSD of the output signal from a fast photodiode when laser light of the horizontal dipole trap is shown onto it. The solid line shows a fit according to Eq.~\eqref{eq:fit-intensity-fluctuations}, yielding an exponent $\tilde{\alpha} = \SI{0.8904(2)}{}$. (b) PSD of the current through the magnetic field coils.}
\label{fig:noise-characterization}
\end{center}
\end{figure}

In order to identify dominant contributions to the experimentally determined noise spectrum $S(\omega)$, we investigate common effects, such as intensity fluctuations of the dipole trap laser and magnetic field fluctuations, that have already been seen to cause homogeneous dephasing in ultracold atomic systems \cite{Kuhr2005, Chow2021}.
The intensity fluctuations are measured by shining the trapping light of the horizontal dipole trap onto a fast photodiode. 
We record the photodiode voltage as a function of time and calculate the power spectral density (PSD) $S_\text{DT}$ of our discrete-time signal. In Fig.~\ref{fig:noise-characterization}(a), $S_\text{DT}$ is visible as the red curve.
The resulting PSD $S_\text{DT}$ comprises two noise components
\begin{equation}
\label{eq:fit-intensity-fluctuations}
	S_\text{DT} (\omega) = \dfrac{S_\text{PLN}}{\omega^{\tilde{\alpha}}}+ S_\text{WN},
\end{equation}
where the first component describes power-law noise with a characteristic, positive exponent $\tilde{\alpha}$ and the second component describes white noise with a constant power spectral density $S_\text{WN}$.
We use Eq.~\eqref{eq:fit-intensity-fluctuations} as a fit model for our data in Fig.~\ref{fig:noise-characterization}(a) and find an exponent $\tilde{\alpha} = \SI{0.8904(2)}{}$. This fit is visible as black solid line in Fig.~\ref{fig:noise-characterization}(a) and agrees very well with $S_\text{DT}$. 
Importantly, the fitted exponent coincides with $\mathit{\alpha} = \SI{0.89(2)}{}$ obtained from the measured visibility evolutions. 
This suggests that the dominant contribution to the dephasing of the qubit originates from noise in the laser intensity which generates the dipole trap. Moreover, previously, we have estimated that some of the dips in the visibility data originate from a resonance of $S(\omega)$ at $\omega_0\approx 2\mathrm{\pi}\times \SI{100}{\hertz}$. Remarkably, a resonance at $\omega_0$ is found in $S_\text{DT}$ and can therefore explain such features in the visibility.

To quantify the magnetic field fluctuations, we cannot directly measure the fluctuations outside of the vacuum system because they differ from the fluctuations at the atoms' position.  
Instead, we record the current through the magnetic field coils as a function of time and calculate the PSD of the signal. The resulting PSD, $S_B$, is shown as brown curve in Fig.~\ref{fig:noise-characterization}(b).  Fundamentally different from $S_\text{DT}$, the PSD $S_B$ is rather flat with no clear decay. 
Such a noise spectrum would result in a very different evolution of the visibility. 
Therefore, we conclude that the background noise floor inferred from the measured coherence evolution is mainly caused by intensity fluctuations of the trapping light and not by magnetic field fluctuations. This finding highlights the magnetic-insensitivity of the two Cs clock states which have been used as stable qubit states.

\section{\label{sec:conclusion-outlook}Conclusion and Outlook}
In conclusion, we have studied the dynamics of a noise-coupled qubit realized in $^{133}$Cs under the effect of dynamical driving. 
Hereby, we have focused on a dynamical decoupling protocol, in particular the CPMG pulse sequence, that has increased the coherence time of the qubit by an order of magnitude. 
Moreover, from the visibility data, we were able to analyze properties of the environmental noise spectrum including its algebraic decay. 
From this, we have calculated analytical estimates for the coherence time which are in good agreement with the measured data. 
To find the origin of the noise affecting the qubit dynamics, we have measured the PSDs determining the noise spectra of the dipole trap laser intensity and the magnetic field. 
With these, we could show that the main noise source originates from the trapping laser intensity while magnetic field fluctuations play only a minor role. 
The latter originates from the magnetic-insensitivity of the used qubit states. 
Remarkably, the algebraic decay obtained from the visibility data is in excellent agreement with the algebraic decay found in the PSD of the dipole trap laser intensity. 
This highlights the ability to measure spectral properties of the environment with a dynamical probe. 

In future work, we will investigate the dynamics of the $^{133}$Cs qubit when it is coupled to a bosonic environment of $^{87}$Rb atoms. 
The qubit dynamics could then be analyzed to potentially measure properties of an interacting environment. 
This is the next step towards the implementation of a single $^{133}$Cs qubit as a versatile, nondestructive quantum probe for noise spectroscopy of engineered atomic baths. 
An important prerequisite for this is that the lifetime of the qubit without the bosonic bath exceeds the lifetime in the presence of the $^{87}$Rb atoms.
By comparing typical lifetimes obtained for the mixture~\cite{Adam2022}, $\tau_{\text{c},\text{Rb}}\approx \SI{1}{\milli\second}$, which are one order of magnitude shorter than the shortest lifetime obtained in this paper, $\tau_\text{c} = \SI{16.2(9)}{\milli\second}$, we conclude that this requirement is fulfilled.
This will enable the distinction of dynamical features emerging from, e.g., noise in the dipole trap laser intensity and features emerging from the bosonic $^{87}$Rb bath.

Besides its potential as a quantum sensor, this setup can also be used to explore the entanglement dynamics of the Cs qubit with the surrounding $^{87}$Rb bath under dynamical driving. 
This creates the possibility to study non-equilibrium phenomena in quantum many-body systems closely connected to spin-boson and polaron physics. 
Additionally, it will be interesting to investigate the possibility of tailoring the qubit-bath coupling with periodic driving which opens the door to quantum-state engineering in open quantum systems.

\begin{acknowledgments}
This work was supported by the Deutsche Forschungsgemeinschaft (DFG, German Research Foundation) via the Collaborative Research Centers SFB/TR185 (Project No. 277625399).
S.B. acknowledges funding by Studienstiftung des deutschen Volkes.
\end{acknowledgments}

\vspace{0.5 cm}

\section*{\label{sec:data-availability}Data availability statement}
The data that support the findings of this study are openly
available at the following URL/DOI: \url{https://doi.org/10.5281/zenodo.7728140}.

\bibliographystyle{apsrev4-2}
\bibliography{manuscript.bib}

\begin{thebibliography}{44}
\bibitem [{Ato()}]{AtomNumberOffset}%
  \BibitemOpen
  \href@noop {} {\bibinfo  {journal} {The measurements shown in the panels (a)
  and (b) of Fig.~\ref{fig:qubit-characterization} exhibit an atom number
  offset of about $\SIrange{6}{7}{}$ Cs atoms. These Cs atoms are remnants of
  the imperfect state preparation process, i.e. they are in states $\ket{F=3,
  m_F \neq 0}$ which do not interact with the microwave but are always imaged.
  For all the other shown measurements, we included an additional series of
  microwave Landau-Zener sweeps and resonant laser pulses before the imaging to
  remove these atoms from the trap, resulting in a typical offset of at most
  two Cs atoms.}\ }\BibitemShut {NoStop}%
\end{thebibliography}

\appendix
\renewcommand{\thefigure}{A\arabic{figure}}
\setcounter{figure}{0}

\section*{\label{app:bfield-calibration}Appendix: Magnetic field calibration}
\begin{figure}
\begin{center}
\includegraphics[width=8.6 cm]{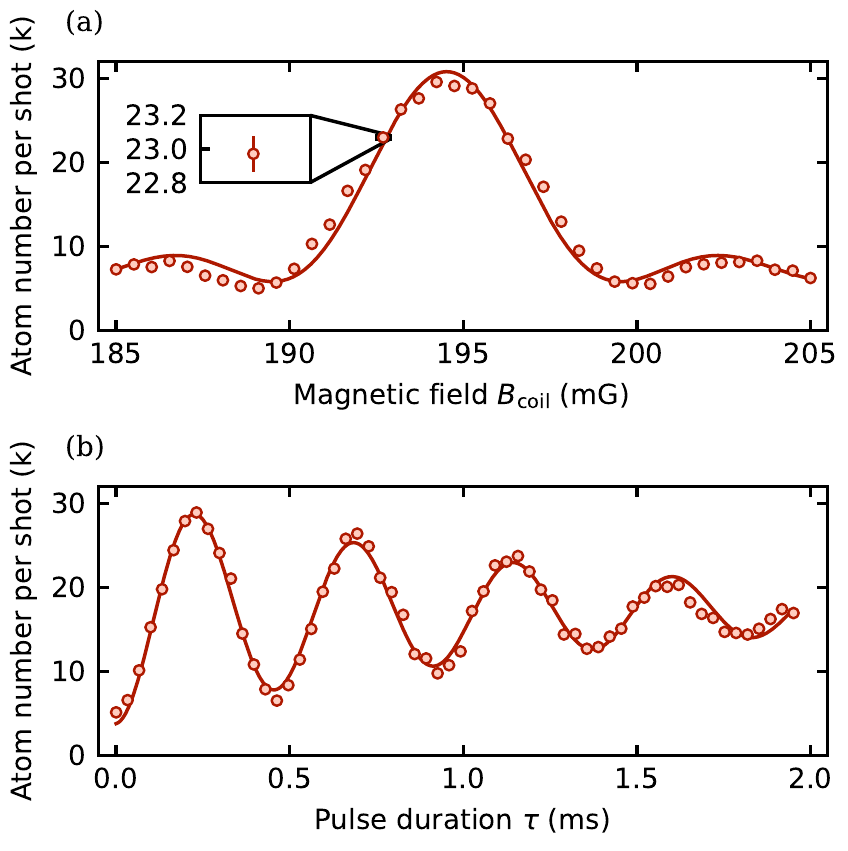}
\caption{\textbf{Magnetic field calibration.} (a)
Microwave spectrum of Rb atoms. The solid line shows a fit according to Eq.~(\ref{eq:mw-spectrum-bfield}) for the extraction of the magnetic field $B_{\text{coil},\text{res}}=\SI{194.52(5)}{\milli\gauss}$. (b) Resonant Rabi oscillations of Rb atoms. Each data point shown in panels (a) and (b) is an average of typically three experimental runs with about $\si{35000}$ atoms each. The inset in panel (a) shows the typical error bar size. All error bars are extracted from the uncertainties in the atom number determination via standard error propagation.}
\label{fig:bfield-calibration}
\end{center}
\end{figure}

We employ microwave spectroscopy on the $\ket{F=1,m_F=0} \rightarrow \ket{F=2,m_F=1}$ transition of the $^{87}$Rb ground state to precisely calibrate the magnetic field $B_\text{exp}$ at the atoms' positions by compensating for the contributions $B_\text{bg}$ of ambient fields.
The Rb atoms are trapped in the second crossed optical dipole trap and are initially prepared in the $\ket{F=1, m_F=0}$ state.
The microwave frequency is fixed to the expected transition frequency given by the Breit-Rabi formula for the desired field magnitude $B_\text{exp}=\SI{198.5}{\milli\gauss}$, neglecting the small differential light shift $\delta_{\text{DT}} = \SI{39}{\hertz}$ of the trapping field.
We vary the externally applied magnetic field $B_\text{coil}$, and thereby also the transition frequency of the Rb atoms, and probe the excited state population by standard absorption imaging using a time-of-flight measurement [Fig.~\ref{fig:bfield-calibration}(a)].
The microwave pulse duration $\tau = \SI{243.9}{\micro\second}$ is chosen such that full population transfer occurs as soon as the resonance condition $B_\text{coil}+B_\text{bg}=B_\text{exp}$ is fulfilled.
The signal follows the typical shape of the spectrum of a Fourier-limited square pulse 
\begin{equation}
N_\text{Rb}(B_\text{coil}) = A \frac{\mathit{\Omega}_{\text{R}}^2}{\mathit{\Omega}^2} \sin^2 \left(\frac{\mathit{\Omega} \mathit{\tau}}{2} \right) + C,
\label{eq:mw-spectrum-bfield}
\end{equation}
where $\mathit{\Omega}^2 = \Delta^2 + \mathit{\Omega}_{\text{R}}^2$, $\Delta = 2 \pi \times (B_\text{coil} - B_{\text{coil},\text{res}}) \times \SI{0.7}{\mega\hertz\per\gauss}$ in the limit of small magnetic fields and $A$  and $C$ are the amplitude and the offset of the fit, respectively.
Notably, the resonant Rabi frequency $\mathit{\Omega}_{\text{R}} = 2 \mathrm{\pi} \times \SI{2.180(5)}{\kilo \hertz}$ is measured independently [Fig.~\ref{fig:bfield-calibration}(b)] so that $A$, $C$, and $B_{\text{coil},\text{res}}$ are the only free parameters of the fit.
\end{document}